\documentclass[floats,floatfix,showpacs,amssymb,prl,twocolumn,superscriptaddress,nofootinbib]{revtex4-1}
\usepackage[utf8]{inputenc}
\usepackage{amsmath}
\usepackage{graphicx}
\usepackage{color}
\usepackage{enumitem}

\newcommand{\cg}[1]{{\color{black}{{#1}}}}

\def\mnras{{Mon. Not. R. Astron. Soc.}}             %
\def\apjl{{Astrophys. J. Lett.}}       
\def\apjs{{Astrophys. J. Suppl.}} 
\def\pasp{{PASP}}             
\def\aap{{Astron. Astrophys.}}  

\begin{document}
\title{Mitigating the counterpart selection effect for standard sirens}

\author{Hsin-Yu Chen} 
\affiliation{Department of Physics, The University of Texas at Austin, 2515 Speedway, Austin, TX 78712, USA}

\author{Colm Talbot} 
\affiliation{Department of Physics and Kavli Institute for Astrophysics and Space Research, Massachusetts Institute of Technology, 77 Massachusetts Ave, Cambridge, MA 02139, USA}

\author{Eve~A. Chase} 
\affiliation{Center for Theoretical Astrophysics, Los Alamos National Laboratory, Los Alamos, NM 87545, USA}

\begin{abstract}
{The disagreement in the Hubble constant measured by different cosmological probes highlights the need for a better understanding of the observations or new physics. The standard siren method, a novel approach using gravitational-wave observations to determine the distance to binary mergers, has great potential to provide an independent measurement of the Hubble constant and shed light on the tension in the next few years.}
To realize this goal, we must thoroughly understand the sources of potential systematic bias {of standard sirens}.
Among the known sources of systematic uncertainties, selection effects originating from electromagnetic counterpart observations of gravitational-wave sources may dominate the measurements {with percent-level bias} and no method to mitigate this effect is currently established. In this Letter, we develop a new formalism to mitigate the counterpart selection effect. We show that our formalism can reduce the systematic uncertainty of standard siren Hubble constant measurement to less than {the statistical uncertainty with a simulated population of 200 observations ($\lesssim 1\%$) for a realistic electromagnetic emission model}. We conclude with how to apply our formalism to different electromagnetic emissions and observing scenarios. 
\end{abstract}
\maketitle

{\em Introduction.--}
The current expansion rate of the Universe, the Hubble constant ($H_0$), is used to determine the age of the Universe. However, the \cg{percent-level} inconsistency between different measurements of this quantity~\citep{planck2018,sh0es22,gaiacollaboration2020gaia,Pesce:2020xfe,Wong:2019kwg,eBOSS:2020yzd} indicates either an insufficient understanding of the Universe or the experiments. Gravitational-wave (GW) standard siren observations offer a promising independent route to resolve this mystery in cosmology~\citep{schutzsiren,holzsiren,2006PhRvD..74f3006D,170817h0,samayah0,hsinyusiren,o3cosmo,2019PhRvL.122f1105F}. 

The distance to a given binary can be determined from GW observations; meanwhile, multiple techniques are available to estimate the redshift of the binaries~\citep{schutzsiren,holzsiren,waltersiren,taylorsiren,messengersiren,farrsiren,josesiren}. With estimates of both distance and redshift, binaries observed in GWs can serve as so-called ``standard sirens'' to measure $H_0$ and other cosmological parameters. In particular, the standard siren \cg{measurements from binary neutron star mergers} have \cg{ideal coverage in redshift ($z\lesssim 0.1$)~\cite{KAGRA:2013rdx,2021CQGra..38e5010C},} well-controlled systematics and are expected to achieve percent-level $H_0$ precision in the coming years, showing great potential to play a critical role in the Hubble tension problem~\citep{samayah0,hsinyusiren,o3cosmo,2019PhRvL.122f1105F}.

However, the standard siren method is not completely systematic-free. Systematic uncertainties associated with the standard siren method have been explored recently~\citep{simonesiren,2020PhRvL.125t1301C,2022arXiv220403614H,2022PhRvD.106d3504M,2020MNRAS.492.3803H,2021A&A...646A..65M,2020MNRAS.495...90N,simonesiren}. The dominant systematic uncertainty was long-thought to be instrumental calibration uncertainty, but a recent study of GW observations showed that calibration is unlikely to be an issue~\cite{2022arXiv220403614H}. \cg{The reconstruction of peculiar velocity fields around the host galaxies can lead to percent-level bias for nearby sources, but the majority of GW events are expected to lie at further distances and are less affected by this bias~\cite{2020MNRAS.492.3803H,2021A&A...646A..65M,2020MNRAS.495...90N}. Other known sources of systematics, e.g. GW waveform accuracy~\cite{LIGOScientific:2018hze,LIGOScientific:2020aai,2020PhRvR...2d3039N} and instrumental non-stationary noise~\cite{2022PhRvD.106d3504M}, are expected to lead to bias at the sub-percent level.} Therefore, \cg{the remaining} percent-level systematic, the electromagnetic (EM) counterpart selection effect~\cite{2020PhRvL.125t1301C}, may be the dominant source of bias.

The identification of EM counterparts of GW sources is the most promising scenario to determine the redshift for standard siren measurements. These ``bright sirens'' allow for precise measurements of the redshift for GW sources. However, unlike GW signals, the luminosity of EM emissions from binary mergers is highly uncertain. For example, although it is known that a short gamma-ray burst and kilonova emission  may accompany a binary neutron star (BNS) merger\cg{~\cite{LIGOScientific:2017ync}}, the angular profile of the EM signals and their {emission mechanisms} are undetermined\cg{~\cite{2019LRR....23....1M,2020ApJ...897..150D,2018NatAs...2..751L}}. Such uncertainty could lead to a selection bias when considering multiple GW-EM events for bright siren $H_0$ analysis~\cite{2020PhRvL.125t1301C}. This bias can be corrected if the EM emission model is known~\cite{2020PhRvL.125t1301C,2023MNRAS.520....1G}. In this Letter, we present a new formalism to mitigate the EM counterpart selection effect when the EM emission model is unknown, a more likely scenario in the near future.

We first describe our formalism, followed by realistic examples assuming the EM emissions are anisotropic. We then demonstrate how to generalize our formalism to different EM emissions and observing scenarios in Discussion. 

{\em Mitigating the counterpart selection effect.--}
In order to measure the Hubble constant to sufficient precision, data from multiple pairs of joint GW-EM observations are combined. If we denote the GW and EM observational data as {$\mathcal{D}=(\mathcal{D}_{\rm GW},\mathcal{D}_{\rm EM})$}, the probability \cg{distribution} of $H_0$ given the data (posterior) can be written using Bayes' theorem~\citep{170817h0,hsinyusiren},
\begin{align}\label{eq:h0pos}
    p(H_0|\vec{\mathcal{D}})&\sim \pi(H_0) \prod_{i}\frac{\int {\cal L}(\vec{\mathcal{D}_i}|\vec\Theta,H_0)p(\vec\Theta|H_0)d\vec\Theta}{\int_{\vec{\mathcal{D}}>\vec{\mathcal{D}_{\rm th}}} {\cal L}(\vec{\mathcal{D}}|\vec\Theta,H_0)p(\vec\Theta|H_0)d\vec\Theta d\vec{\mathcal{D}}},
\end{align}
where $\vec{\mathcal{D}}=(\vec{\mathcal{D}_1}, \vec{\mathcal{D}_2}, \vec{\mathcal{D}_3} ...)$ {includes} the data from events $i=1,2,3...$, $\vec{\mathcal{D}_{\rm th}}$ denotes the detection threshold of detectors \cg{(e.g., the minimum signal-to-noise ratio in each GW detector, the limiting magnitude for each filter of the telescope, the minimum number of positive detection over certain EM follow-up cadence, etc.)}, $\pi(H_0)$ is the prior on $H_0$, and $\vec\Theta$ is a collection of binary physical parameters, such as luminosity distance, redshift, \cg{inclination angle,} and mass. {We fix other cosmological parameters to the Planck values assuming flat-$\Lambda$CDM cosmology~\cite{planck2018}, while they can also be inferred by replacing $H_0$ with \cg{a set of cosmological parameters of interest,} $\Omega_{\rm cosmo}=(H_0,\Omega_m,\Omega_{\Lambda}...)$\cg{,} throughout this Letter.}

The event likelihood ${\cal L}(\mathcal{D}_i|\vec\Theta,H_0)$ is the \cg{value of the probability density function of possible data evaluated at the observed} data $\mathcal{D}_i$ given the binary physical parameters $\vec\Theta$ and the Hubble constant $H_0$\cg{~\cite{170817h0,hsinyusiren}}. This \cg{likelihood} depends on the GW and EM emission models of the binary as well as the sensitivities of the GW and EM observatories. We write the dependency on GW and EM emission models and observatory sensitivities explicitly as ${\cal L}(\mathcal{D}_i|\vec\Theta,H_0,\vec\alpha,\vec\beta)$, where $\vec\alpha$ describes the GW emission model and observatory sensitivity, and $\vec\beta$ \cg{describes the EM emission model and sensitivity}. 

How the binary physical parameters $\vec\Theta$ affect the GW luminosity is well-understood from general relativity. With known GW observatory sensitivities, the detectability of GW signals from binary with physical parameters $\vec\Theta$ can be estimated, and therefore we assume $\vec\alpha$ is known. This is not the case for EM counterpart emission. Due to the uncertainty of EM emission models, even if the EM observatory sensitivity is known, $\vec\beta$ is not known. If the effect of $\vec\beta$ is ignored, as has been done for previous bright siren analyses (e.g., ~\cite{170817h0,2019NatAs...3..940H}), the $H_0$ estimate can be biased. This is known as the \textit{counterpart selection effect}. 

\textbf{We mitigate this by estimating $H_0$ and $\vec\beta$ simultaneously. }To do so, \textbf{we analyze all BNS events regardless of whether they have observable EM counterparts.} As in Eq.~\ref{eq:h0pos}, we write the posterior as
\begin{align}\label{eq:h0betapos}
    \nonumber p(H_0,\vec\beta|\vec{\mathcal{D}}, \vec{\alpha})&\sim \pi(H_0,\vec\beta) \times \\ &\prod_{i}\frac{\int {\cal L}(\vec{\mathcal{D}_i}|\vec\Theta,H_0,\vec\alpha,\vec\beta)p(\vec\Theta|H_0)d\vec\Theta}{\int_{\vec{\mathcal{D}}>\vec{\mathcal{D}_{\rm th}}} {\cal L}(\vec{\mathcal{D}}|\vec\Theta,H_0,\vec\alpha,\vec\beta)p(\vec\Theta|H_0)d\vec\Theta d\vec{\mathcal{D}}}.
\end{align}
Note that $\vec\alpha$ is assumed to be known. Also, $\vec\alpha$ and $\vec\beta$ do not affect the intrinsic distribution of the binary physical parameters $p(\vec\Theta|H_0)$.

Since EM counterparts \cg{used in standard siren analysis will mostly be found by followup of GW detections}, the events we use for the analysis are solely determined by the GW selection function. I.e., the integral over $\mathcal{D}$ in the denominator of Eq.~\ref{eq:h0betapos} includes events with and without observable EM counterparts, regardless of their EM observabilities. The denominator of Eq.~\ref{eq:h0betapos} in the product is then
\begin{align*}
    &\int_{\vec{\mathcal{D}}>\vec{\mathcal{D}_{\rm th}}} {\cal L}(\mathcal{D}|\vec\Theta,H_0,\vec\alpha,\vec\beta)p(\vec\Theta|H_0)d\vec\Theta d\mathcal{D}\\
    &=\int_{\vec{\mathcal{D}}_{\rm GW}>\vec{\mathcal{D}}_{\rm GW,th},\vec{\mathcal{D}}_{\rm EM}>\vec{\mathcal{D}}_{\rm EM,th}} {\cal L}(\vec{\mathcal{D}}|\vec\Theta,H_0,\vec\alpha,\vec\beta)p(\vec\Theta|H_0)d\vec\Theta d\vec{\mathcal{D}}\\ 
    &+\int_{\vec{\mathcal{D}}_{\rm GW}>\vec{\mathcal{D}}_{\rm GW,th},\vec{\mathcal{D}}_{\rm EM}\leq\vec{\mathcal{D}}_{\rm EM,th}}{\cal L}(\vec{\mathcal{D}}|\vec\Theta,H_0,\vec\alpha,\vec\beta) p(\vec\Theta|H_0)d\vec\Theta d\vec{\mathcal{D}} \\
    &=\int_{\vec{\mathcal{D}}_{\rm GW}>\vec{\mathcal{D}}_{\rm GW,th}} {\cal L}(\vec{\mathcal{D}}|\vec\Theta,H_0,\vec\alpha)p(\vec\Theta|H_0)d\vec\Theta d\vec{\mathcal{D}}. 
\end{align*}
\textit{This is the same as that in {existing} bright siren inference when no EM selection is in effect. }

In the following, we demonstrate our method using a simple model for counterpart selection effects due to anisotropic EM emission. We will demonstrate other scenarios in Discussion. 

If the EM emission is anisotropic, the inclination angle of a binary $\iota$ will affect the furthest luminosity distance $D_L$ at which counterparts can be observed. For a given telescope configuration (filter, exposure time, camera, etc.), we define the telescope's maximum observable luminosity distance of the merger as $\epsilon(\iota,\vec\beta)$. 

We first explicitly separate \cg{$\vec\Theta$} into relevant physical parameters ($D_L$, $\iota$) from other \cg{physical } parameters $\vec{\Theta}^{'}$ ( \cg{$\vec{\Theta}\supset \{D_L,\iota,\vec{\Theta^{'}}\}$}) and recognize that the distance is uniquely determined by $z$ and $H_0$, ${\cal L}(\mathcal{D}|\vec\Theta,H_0,\vec\alpha,\vec\beta) = {\cal L}(\mathcal{D}|D_L(z,H_0),\iota,\vec\Theta^{'},\vec\alpha,\vec\beta)$. Since counterparts can only be found for $D_{L} \leq \epsilon(\iota,\beta)$, we distinguish between cases where we can and cannot identify a counterpart,
\begin{widetext}
\begin{align}\label{eq:heaviside}
{\cal L}(\mathcal{D}|D_L(z,H_0),\iota,\vec\Theta^{'},\vec\alpha,\vec\beta)= \left\lbrace
\begin{array}{r@{}l}
&{\cal L}(\mathcal{D}_{\rm GW}, \mathcal{D}_{\rm EM}|D_L(z,H_0),\iota,\vec\Theta^{'},\vec\alpha)\mathcal{H}[\epsilon(\iota,\vec\beta)-D_L(z,H_0)]\qquad{({\rm Counterpart})}\\
&{\cal L}(\mathcal{D}_{\rm GW}|D_L(z,H_0),\iota,\vec\Theta^{'},\vec\alpha)\mathcal{H}[D_L(z,H_0)-\epsilon(\iota,\vec\beta)].\qquad\hfill{({\rm No\,\, counterpart})}
\end{array}\right.
\end{align} 
\end{widetext}
The quantity $\mathcal{H}$ is the Heaviside function\footnote{In practice EM instrumental noise will lead to a different functional form for the selection function. The Heaviside function can be trivially replaced by an appropriate sigmoid function.}. This formalism allows for the use of events with and without EM counterparts. Even if no counterpart is identified, the luminosity distance and inclination angle space are constrained.

{\em Application to \cg{simulated} observations.--}
To demonstrate our method, we follow the method in~\cite{2019PhRvX...9c1028C} to simulate 200 1.4-1.4$M_{\odot}$ BNS detected by LIGO-Hanford, LIGO-Livingston, and Virgo operating at the proposed fourth observing run (O4) sensitivity~\citep{KAGRA:2013rdx,LIGOScientific:2014pky,VIRGO:2014yos}. The GW detection threshold is set at a network signal-to-noise ratio of 12\cg{~\cite{KAGRA:2013rdx,2021CQGra..38e5010C}}. We use the Bayesian algorithm developed in~\cite{2019PhRvX...9c1028C} to estimate the line-of-sight luminosity distance-inclination angle posterior when EM counterparts are found and the sky direction of the GW sources is determined, and \texttt{Bilby}~\citep{2019ApJS..241...27A,2020MNRAS.499.3295R,2021MNRAS.507.2037A} for full parameter estimation when the events are dark.

We consider two kilonova emission models. 
Both models are selected from a grid of 900 two-dimensional axisymmetric radiative transfer simulations from~\cite{2021ApJ...918...10W}\cg{, which spans the full range of anticipated kilonova properties}.
\cg{All simulated models are} rendered at 54 inclination angles, each subtending an equal solid angle from 0$^{\circ}$ to 180$^{\circ}$ (see~\cite{2021ApJ...918...10W} for more details).
\cg{Of these 900 simulations, we select two simulated models with significantly different angular profiles.
Model 1 displays substaintial angular variation, typical of kilonova emission, with face-on inclinations resulting in brighter emission.
For more ``realistic'' kilonovae, we selected dyanamical and wind ejecta component masses that result in an $r$-process abundance pattern similar to the neutron-capture elements observed in the Solar System~\cite{2022arXiv220602273R}. 
All other model properites were selected to result in maximal angular variation.
Specifically,} Model 1 represents a kilonova with a toroidal dynamical ejecta component with a mass of 0.01~$M_{\odot}$ and mean velocity of 0.3$c$ in addition to a ``peanut-shaped'', high-$Y_e$ wind ejecta component with a mass of 0.03~$M_{\odot}$ and mean velocity of 0.15$c$.

For the second kilonova model, we intentionally chose a model that has minimal dependence on the inclination angle in order to test the performance of our method when the EM emission is isotropic.
\cg{We selected the kilonova with the least angular variation of all 900 simulated models in~\cite{2021ApJ...918...10W}.}
Model 2 represents a kilonova with a toroidal dynamical ejecta component with a mass of 0.001~$M_{\odot}$ and mean velocity of 0.05$c$ in addition to a spherical, low-$Y_e$ wind ejecta component with a mass of 0.1~$M_{\odot}$ and mean velocity of 0.3$c$.
\cg{This model was also used as an example of minimal angular variation in Fig. 3 of~\cite{2021ApJ...918...10W}.}

\begin{figure}
    \centering
    \includegraphics[width=1.0\linewidth]{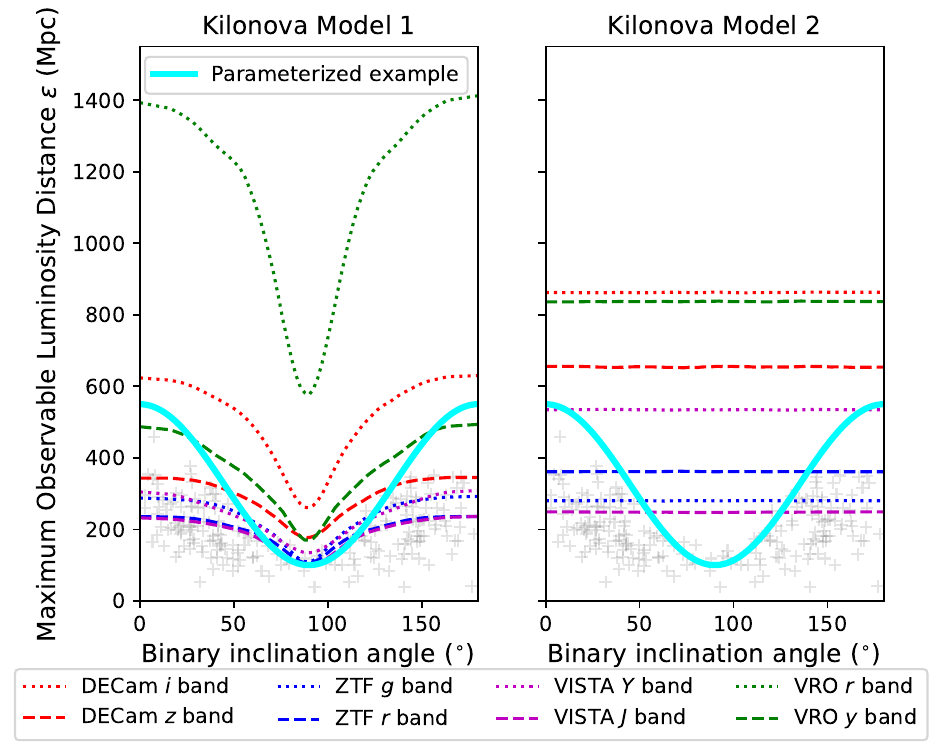}
    \caption{Maximum observable luminosity distance as a function of the binary inclination angle for the two kilonova models and different telescope configurations (listed in Table~\ref{tab:telescope}) we consider. The gray crosses mark the inclination angle and luminosity distance of our simulated BNS detections. We also give an example of the function (Eq.~\ref{eq:betafunc}) we used to parametrize the shape of curves in this figure (cyan line, $\beta_1=450, \beta_2=100$ in this example).}
    \label{fig:maxobsDL}
\end{figure}

Next, we explore the detectability of the kilonovae at different inclination angles for a variety of optical/near-infrared instruments and filters (Table~\ref{tab:telescope}). 
Limiting magnitudes are based on telescope designed sensitivities or from performance in LIGO-Virgo third observing run, as detailed in~\cite{2022ApJ...927..163C}.
The time of observation after the merger and the exposure time assumed are also listed.
We note that this represents a selection of possible limiting magnitudes, and a wide range of instrument sensitivity is possible with different exposure times, background contributions, and environment factors.
Our choice of the telescope configurations is not intended to compare the performance of the instruments, but instead represents a wide variety of observing scenarios. 
We follow~\cite{2022ApJ...927..163C} to combine simulated observer-frame spectroscopic kilonova emission with an instrument's bandpass filter function for a variety of luminosity distances.
By comparing to the limiting magnitude of the selected telescope configurations, we are able to estimate the maximum observable luminosity distance for the kilonova as a function of binary inclination angle, at a given observation time.
In Fig.~\ref{fig:maxobsDL} we show how the maximum observable luminosity distance depends on the binary inclination angle with dashed and dotted curves. 
As expected, the results for kilonova model 1 are significantly more sensitive to the inclination angle than model 2.

\begin{table}[]
    \centering
    \begin{tabular}{c|c|c|c|c|c}
        Instrument & Filter & Exp. Time (s) & $m_{\rm lim}$& Obs. Time (hrs) & Ref. \\ \hline
        DECam & \textit{i} & 90 & 22.5 & 18 & \citep{2016ApJ...823L..33S} \\
        DECam & \textit{z} & 90 & 21.8 & 12 & \citep{2016ApJ...823L..33S}\\
        ZTF & \textit{g} & 30 & 20.8 & 12 & \citep{2019PASP..131a8002B}\\
        ZTF & \textit{r} & 30 & 20.6 & 12 & \citep{2019PASP..131a8002B} \\
        VISTA & \textit{Y} & 360 & 21.5 & 24 & \cite{2013Msngr.154...35M,2015MNRAS.446.2523B} \\
        VISTA & \textit{J} & 360 & 21.0 & 48 & \cite{2013Msngr.154...35M,2015MNRAS.446.2523B}\\
        VRO & \textit{r} & 30 & 24.2 & 12 & \citep{2019ApJ...873..111I}\\
        VRO & \textit{y} & 30 & 22.3 & 24 & \citep{2019ApJ...873..111I}
    \end{tabular}
    \caption{The instruments, filters, exposure time, limiting magnitude (AB) and observing time after merger we explored in this paper. We employ typical exposure times and limiting magnitudes, as detailed in \cite{2022ApJ...927..163C}, but note that these observing conditions are highly variable.}
    \label{tab:telescope}
\end{table}

We then parametrize the maximum observable distance as
\begin{equation}\label{eq:betafunc}
    \epsilon(\iota,\vec\beta)=\beta_1 {\rm cos}^2(\iota)+\beta_2 \,\, {\rm Mpc},
\end{equation}
and show an example of this model in cyan in Fig.~\ref{fig:maxobsDL}. \cg{This is a generic functional form that describes a lower observable distance when the binary is edge-on, a fairly common feature among kilonova simulations. We note that more refined models may be adopted when larger observed populations are available.}

For each of our observing scenarios, we determine which of the 200 simulated BNS signals would have identifiable counterparts based on their luminosity distance and inclination angle.
We then use \texttt{emcee}~\citep{2013PASP..125..306F} to infer $(H_0,\beta_1,\beta_2)$.

{\em Result.--}
Among the two kilonova models and eight telescope configurations we consider, some of the configurations are able to find the kilonovae for all BNSs we simulated, so we do not expect counterpart selection effect in $H_0$ inference. We skip these in the presentation of results. 

\begin{figure}
    \centering
    \includegraphics[width=1.0\linewidth]{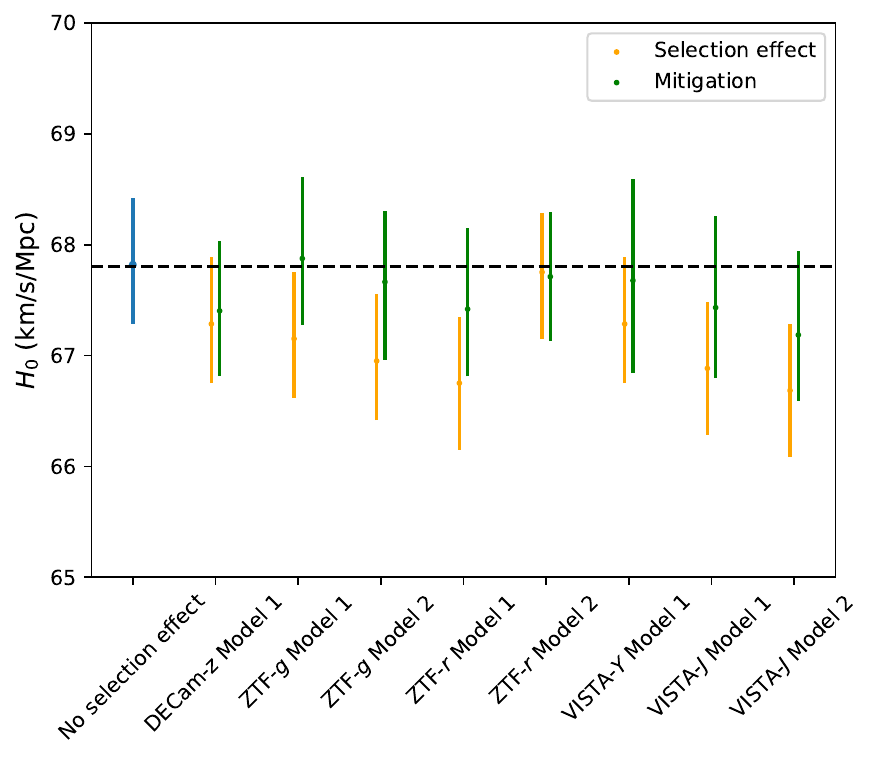}
    \caption{The median and symmetric 68\% credible interval for $H_0$ inferred using 200 simulated BNS detections. The horizontal axis labels the telescope configurations (c.f. Table~\ref{tab:telescope}) and kilonova models (c.f. Fig.~\ref{fig:maxobsDL}). \textit{Selection effect} (orange)- When $H_0$ inference ignores the presence of counterpart selection effect. \textit{Mitigation} (green)-$H_0$ inference using the formalism we present in Eq.~\ref{eq:h0betapos}. For comparison, we also show the $H_0$ inferred when all counterparts are captured and there is no counterpart selection effect (blue). The horizontal dashed line denotes the $H_0$ we picked for the simulations.}
    \label{fig:mitigation}
\end{figure}

In Fig.~\ref{fig:mitigation} we show the inferred median and 68\% symmetric credible interval for $H_0$ for the remaining configurations with (green) and without (orange) correcting for the counterpart selection effect. {The results have marginalized over $\vec\beta$.}
In blue we show the results when all BNS have an identified EM counterpart.
When not correcting for selection effects, we see up to 2\% bias in $H_0$ in our simulations. Our formalism \cg{reduces the systematic bias to less than the statistical uncertainty ($\sim 1\%$) in all cases for our simulated observations (the true value is contained within the $68\%$ credible interval). Additionally,} missing some of the EM counterparts naturally leads to larger $H_0$ measurement uncertainties. In our simulations, we find 3 -- 50\% larger uncertainties when counterpart selection effects are present. 

We note that to elucidate the origin of Hubble tension, we will likely need fewer than 200 BNSs\cg{~\cite{samayah0,hsinyusiren,o3cosmo,2019PhRvL.122f1105F}} and it is unlikely to observe all of them in O4 as in our simulations~\citep{gwtc3,gwtc3pop}. We combined more events than needed in order to avoid statistical fluctuation in the demonstration and to reveal any underlying systematic uncertainty. However, our formalism can be applied to other GW detector sensitivities and network configurations.

{\em Discussion.--}
In this Letter, we present a new bright siren inference formalism that mitigates the EM counterpart selection effect. By including GW events both with and without counterparts, \cg{our formalism successfully mitigates the systematic bias introduced by the counterpart selection effect for a simulated population of 200 observations}. 
{The method \cg{does not require precise assumptions of the} EM models, avoiding additional modeling systematics.} \cg{It is possible there are remaining systematics below the statistical uncertainty} due to the simplified functional form employed to parametrize the maximum observable luminosity distance (Eq.~\ref{eq:betafunc}). Future developments in observations, theories, and numerical methods will further reduce \cg{any such} bias.

Even with the nearly isotropic kilonova model (Model 2), we see 2\% bias in $H_0$ for some telescope configurations. 
The bias in these scenarios \cg{mainly originates from the limitation of EM instrument sensitivities, as some face-on binaries lie beyond the instrument's maximum observable distance (Fig.~\ref{fig:maxobsDL}, right panel). Still,} our formalism is able to mitigate the bias. 
\cg{We note that in this case, the extra degree of freedom allowed by our nearly agnostic prior on $\beta_1$ ($\beta_1\in[0,1000]$Mpc) leads to the median recovered $H_{0}$ value falling below the truth. Since the correct model is contained within the prior distribution, we can safely expect the true value to continue to be contained within the statistical uncertainty as more events are observed.}
{In the future, by comparing different EM observations and inclination angle of the events, it will become clear whether kilonovae are better modeled as having isotropic emission, and we can adopt a constant maximum observable distance model (i.e., \cg{$\beta_1=0$} in Eq.~\ref{eq:betafunc}). If so, we find that the $H_0$ bias becomes less than 0.35\% with our formalism.}

Although we demonstrate our formalism with the selection effect associated with anisotropic EM emission observed by a selected set of instrument configurations, the formalism can easily be generalized to many other scenarios including:
\begin{itemize}[leftmargin=*]
    \item \textit{Mass-dependent EM emission}: The EM emission from BNSs and neutron star-black hole mergers strongly depends on the component masses of the binaries\cg{~\cite{dyn,rem}} and can lead to a similar counterpart selection effect in bright siren inference. In order to mitigate this effect, one could choose an appropriate functional form $\epsilon(m_1,m_2,\vec\beta)$. Since the component masses of binaries are better measured in GWs than the inclination angle, the mass-associated counterpart selection effect {could} be easier to mitigate.
    \item \textit{Multiple instruments and/or different observing strategies}: The counterparts of GW events will likely be discovered by different instruments or observing strategies (e.g., searching for counterparts at different depths, colors, and time after mergers.) Each variation leads to a different counterpart selection effect. We can take $\vec\beta={\vec\beta_a,\vec\beta_b,\vec\beta_c...}$ to represent each variation. If the EM observations are random to GW events, i.e. the search instruments and strategies do not depend on binary's physical parameters, the posterior for $H_0$ can be written as
    \begin{equation}\label{eq:multi}
        p(H_0|\{\vec{\mathcal{D}_a},\vec{\mathcal{D}_b},\vec{\mathcal{D}_c}...\})\sim \pi(H_0) \prod_{k=a,b,c...} {\cal L}(\vec{\mathcal{D}_k} | H_0),
    \end{equation}
    where ${\cal L}(\vec{\mathcal{D}_k} | H_0)$ is the product of the likelihoods for each event observed with scenario $k$ marginalized over $\vec\beta_{k}$.
    If the choice of instruments or strategies depends on the physical parameters of the GW signals, one can account for this as an additional \textit{known} selection effect. 
    \item \textit{Partial sky coverage}: The large GW sky localization area can make it difficult for a single instrument to cover the entire localization map. If so, the integral over $\vec\Theta$ in the numerator of Eq.~\ref{eq:h0betapos} has to be limited to the sky directions $(\vec{\Omega_A},\vec{\Omega_B},\vec{\Omega_C}...)$ covered by instruments (A,B,C...), respectively.
    \item \textit{No counterpart due to other reasons}: Non-detection of counterparts due to weather, instrument availability, solar position etc., are likely to be random, and so the associated GW events can be removed from the bright siren inference without leading to a selection bias.  
    \item \textit{Hostless counterpart}: A counterpart \cg{may} be found without an associated host for the determination of redshift.
    {However,} the observation of a counterpart indicates that the event lies within the maximum observable luminosity distance. One can use the counterpart-found scenario {(e.g., $\mathcal{H}[D_L(z,H_0)-\epsilon(\iota,,\vec\beta)]$ case in Eq.~\ref{eq:heaviside})}
    and integrate over possible redshifts in the numerator of Eq.~\ref{eq:h0betapos}.  
\end{itemize}

{As LIGO-Virgo-KAGRA continue to observe with improved sensitivities, joint GW-EM detections will push toward percent-level $H_0$ measurements in the next few years. Our new formalism mitigates a dominant source of systematic uncertainty with minimal dependence on EM modeling for the measurement, ensuring a reliable path toward resolving the $H_0$ tension.}


\section*{Acknowledgements}
The authors would like to thank Simone Mastrogiovanni for their review and suggestions to this work. The authors would like to thank Maya Fishbach, Jonathan Gair, Daniel Holz, Will Farr, Marko Ristic, and Aaron Zimmerman for useful discussions.
HYC is supported by the National Science Foundation under Grant PHY-2308752. 
HYC was supported by NASA through NASA Hubble Fellowship grant No.\ HST-HF2-51452.001-A awarded by the Space Telescope Science Institute, which is operated by the Association of Universities for Research in Astronomy, Inc., for NASA, under contract NAS5-26555.
CT is supported by an MIT Kavli Fellowship.
This analysis was made possible by the {\tt numpy}~\citep{numpy,numpy:2020}, {\tt SciPy}~\citep{Virtanen:2019joe}, {\tt matplotlib}~\citep{Hunter:2007ouj}, {\tt emcee}~\citep{ForemanMackey:2012ig,2013PASP..125..306F}, {\tt pandas}~\citep{reback2020pandas, mckinney-proc-scipy-2010}, and {\tt astropy}~\citep{Robitaille:2013mpa, Price-Whelan:2018hus} software packages. 
This material is based upon work supported by NSF's LIGO Laboratory which is a major facility fully funded by the National Science Foundation. This is LIGO Document Number LIGO-P2300222.

\bibliography{ref}

\end{document}